\documentclass[twocolumn,showpacs,aps,prb,superscriptaddress,floatfix]{revtex4}
\usepackage{graphicx}
\usepackage{dcolumn}
\usepackage{bm}

\usepackage{color}

\begin{document}

\preprint{}

\title{Dimensional Evolution of Spin Correlations in the Magnetic Pyrochlore Yb$_2$Ti$_2$O$_7$}

\author{K.A. Ross}
\affiliation{Department of Physics and Astronomy, McMaster University,
Hamilton, Ontario, L8S 4M1, Canada}
\author{L.R. Yaraskavitch}
\affiliation{Department of Physics and Astronomy and Guelph-Waterloo Physics Institute, University of Waterloo, Waterloo, ON, N2L 3G1, Canada}
\affiliation{Institute for Quantum Computing, University of Waterloo, Waterloo, ON, N2L 3G1, Canada}
\author{M. Laver} 
\affiliation{Materials Research Division, Ris\"{o} DTU, Technical University of Denmark, DK-4000 Roskilde, Denmark}
\affiliation{Nano-Science Center, Niels Bohr Institute, University of Copenhagen, DK-2100 K\"{o}benhavn, Denmark}
\affiliation{Laboratory for Neutron Scattering, Paul Scherrer Institut, 5232 Villigen PSI, Switzerland}
\author{J.S. Gardner} 
\affiliation{Indiana University, 2401 Milo B. Sampson Lane, Bloomington, Indiana 47408, USA} 
\affiliation{National Institute of Standards and Technology, 100 Bureau Drive, MS 6102, Gaithersburg, Maryland 20899-6102, USA}
\author{J. A. Quilliam}
\affiliation{Department of Physics and Astronomy and Guelph-Waterloo Physics Institute, University of Waterloo, Waterloo, ON, N2L 3G1, Canada}
\affiliation{Institute for Quantum Computing, University of Waterloo, Waterloo, ON, N2L 3G1, Canada}
\author{S. Meng}
\affiliation{Department of Physics and Astronomy and Guelph-Waterloo Physics Institute, University of Waterloo, Waterloo, ON, N2L 3G1, Canada}
\affiliation{Institute for Quantum Computing, University of Waterloo, Waterloo, ON, N2L 3G1, Canada}
\author{J.B. Kycia}
\affiliation{Department of Physics and Astronomy and Guelph-Waterloo Physics Institute, University of Waterloo, Waterloo, ON, N2L 3G1, Canada}
\affiliation{Institute for Quantum Computing, University of Waterloo, Waterloo, ON, N2L 3G1, Canada}
\author{D. K. Singh} 
\affiliation{National Institute of Standards and Technology, 100 Bureau Drive, MS 6102, Gaithersburg, Maryland 20899-6102, USA}
\affiliation{Dept. of Materials Science and Engineering, University of Maryland, College Park, MD 20742}
\author{Th. Proffen} 
\affiliation{Los Alamos National Laboratory, Los Alamos, New Mexico 87545, USA}
\affiliation{Spallation Neutron Source, Experimental Facilities Division, Oak Ridge National Laboratory, P.O. Box 2008, Oak Ridge, TN 37831, USA}
\author{H.A. Dabkowska} 
\affiliation{Brockhouse Institute for Materials Research, McMaster University, Hamilton, Ontario, L8S 4M1, Canada}

\author{B.D. Gaulin} 
\affiliation{Department of Physics and Astronomy, McMaster University,
Hamilton, Ontario, L8S 4M1, Canada}
\affiliation{Brockhouse Institute for Materials Research, McMaster University, Hamilton, Ontario, L8S 4M1, Canada}
\affiliation{Canadian Institute for Advanced Research, 180 Dundas St.\ W.,Toronto, Ontario, M5G 1Z8, Canada}


\bibliographystyle{prsty}

\begin{abstract}
The pyrochlore material Yb$_2$Ti$_2$O$_7$ displays unexpected quasi-two-dimensional (2D) magnetic correlations within a cubic lattice environment at low temperatures, before entering an exotic disordered ground state below T=265mK.   We report neutron scattering measurements of the thermal evolution of the 2D spin correlations in space and time.  Short range three dimensional (3D) spin correlations develop below 400 mK, accompanied by a suppression in the quasi-elastic (QE) scattering below $\sim$ 0.2 meV.  These show a slowly fluctuating ground state with spins correlated over short distances within a kagome-triangular-kagome (KTK) stack along [111], which evolves to isolated kagome spin-stars at higher temperatures.  Furthermore, low-temperature specific heat results indicate a sample dependence to the putative transition temperature that is bounded by 265mK, which we discuss in the context of recent mean field theoretical analysis.  

\end{abstract}


\maketitle


\section{\label{sec:level1}Introduction}
The corner-sharing tetrahedral geometry of the pyrochlore lattice favors strong \textit{geometric frustration}, wherein a competition of exchange interactions between localized magnetic moments results directly from their arrangement within the lattice.  This phenomenon tends to suppress transitions to magnetic long range order (LRO), leaving in their place unusual short-range magnetic correlations at low temperatures.  The form of such correlations varies between frustrated systems, with ground states typically determined by subleading terms in the relevant Hamiltonian, such as dipole interactions, further neighbor exchange, and structural distortions within the underlying crystal lattice.\cite{lacroix2011introduction, gardner2010magnetic}  The rare earth titanate series of materials, which have the chemical formula \textit{R}$_2$Ti$_2$O$_7$ with \textit{R} being a trivalent magnetic rare earth ion, crystallize into the face centered cubic (fcc) space group Fd$\bar3$m which characterizes the pyrochlore lattice.  This series provides excellent examples of the variety of exotic ground states that can emerge in nearly identical chemical environments when geometric frustration is at play.  Several of the rare earth titanates exhibit disordered ground states which persist down to 30mK, namely the spin ices Dy$_2$Ti$_2$O$_7$ and Ho$_2$Ti$_2$O$_7$, \cite{bramwellSpinIce, bram_ging, clancy2009revisiting} and the enigmatic spin liquid Tb$_2$Ti$_2$O$_7$.\cite{gardner_spinliq, gardner2001neutron}  Others in the series do order magnetically, such as Gd$_2$Ti$_2$O$_7$\cite{Raju1999gdtio, Champion2001gdtio,stewart2004phase} and Er$_2$Ti$_2$O$_7$,\cite{champion2003, ruff} but the nature of the ordering is unusual in both cases.   In all of these cases, the source of the frustration (or lack thereof) is well-understood in terms of the spin interactions and the single-ion anisotropy imposed by the crystal electric field.  In particular, the spin ices combine effective ferromagnetic (FM) exchange with strong Ising-like anisotropy, such that the moments are constrained to point into or out of each tetrahedron.  

\begin{figure}[!htb]  
\centering
\includegraphics[ angle=90, width=8.5cm]{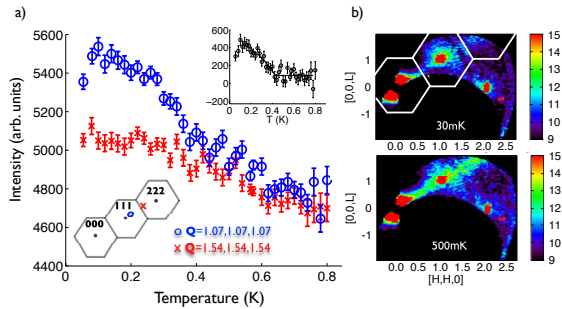}
\caption{(color online) a) The temperature dependence of neutron scattering intensity at two \textbf{Q} points that lie on the rod of scattering, \textbf{Q}$_{1.54}$ (red x's) and \textbf{Q}$_{1.07}$ (blue o's) (10K background subtracted). Bottom Inset: the position of the two \textbf{Q} points is shown relative to the fcc Brillouin zone boundaries in the (HHL) plane.  Top Inset: the difference of intensities, I$_{1.07}$ - I$_{1.54}$, with the solid line as a guide to the eye. b)  Elastic time-of-flight neutron scattering data at 30mK and 500mK (reproduced from Ref. \onlinecite{ross2009}).  Error bars represent $\pm$ 1$\sigma$.}
\label{fig:fig1}
\end{figure}

Yb$_2$Ti$_2$O$_7$ provides an intriguing contrast to the spin ices. It is known to combine FM exchange ($\theta_{CW}$$\sim$600mK to 800mK\cite{bramwell_mag, hodgescrysfield}) with XY anisotropy.\cite{hodgesfluc, malkin, cao2009aniso}  The continuous degree of freedom provided by the XY anisotropy leads one to naively expect an ordered FM ground state.  However, despite what appears to be a transition near 200mK,\cite{blote, hodgesfluc, dalmas2006studies} most experimental evidence suggests that Yb$_2$Ti$_2$O$_7$ displays a disordered ground state down to 30mK in zero-magnetic field.\cite{ross2009, hodgesfluc}  Application of a modest magnetic field along [110] at 30 mK induces a transition to a polarized, ordered phase with well defined spin wave excitations.\cite{ross2009} Using anisotropic exchange Hamiltonians, recent theoretical studies have sought to understand the spin wave dispersions in the magnetic field induced ordered state,\cite{leonspinwaves} as well as the diffuse scattering at relatively high temperatures in zero field \cite{thompson2011rods} and the local spin susceptibility.\cite{cao2009aniso, thompson2011local} Interestingly, some of these studies indicate that the dominant exchange interaction is the Ising component, which may explain the lack of simple FM behaviour in Yb$_2$Ti$_2$O$_7$, making it an \emph{exchange} analog to the spin ices.\cite{leonspinwaves, cao2009aniso}

An early heat capacity (C$_p$) measurement showed an anomaly at T=214mK, indicating a possible phase transition, in addition to a broad Schottky-like hump at 2K.\cite{blote} The putative transition, whose temperature varies somewhat in the literature and which we henceforth label T$_c$, has previously been explored through several techniques.\cite{hodgesfluc,gardner,yasui,ross2009}  Much of this characterization is consistent with the absence of conventional LRO below T$_c$, although single crystal neutron diffraction and AC susceptibility results presented in Ref. \onlinecite{yasui}, suggest a collinear FM ground state.  Hodges \emph{et al} found a discontinuous change in the spin fluctuation frequency at T$_c$=240mK as measured by M\"ossbauer and $\mu$SR, along with an absence of magnetic Bragg peaks in neutron powder diffraction.\cite{hodgesfluc}  The specific heat of the sample used in Ref. \onlinecite{hodgesfluc} was found to display a sharp anomaly at 250mK,\cite{dalmas2006studies} close to the observed first order drop in the spin fluctuation rate.  Single crystal neutron scattering studies revealed a pattern of diffuse scattering, notably taking the form of \emph{rods} of scattering along the $<$111$>$ directions, with a characteristic QE energy scale of 0.3meV.\cite{ross2009, bonvillehyp} This diffuse scattering, which indicates short range spin correlations, is present both above and below T$_c$, but a qualitative difference was observed between 500mK and 30mK that indicated short-range 3D spin correlations at the lowest temperatures.\cite{ross2009}   The observed change was that the rod of scattering became more strongly peaked near the (111) zone center, moving from a featureless rod at 500mK to a rod with structure at 30mK.  The conventional interpretation of a \emph{flat} rod of scattering in any diffraction pattern is the presence of 2D correlations.  Though the rod of scattering is flat at 500mK, this interpretation is striking in the present case, as it implies a magnetic decomposition of the 3D pyrochlore structure into a stacking of alternating kagome and triangular planes along $<$111$>$, breaking the underlying cubic symmetry.

It should be noted that in Ref. \onlinecite{thompson2011rods}, where exchange parameters were extracted by fitting a modified measurement of the diffuse scattering pattern in the HHL plane at 1.4K , the real space spin correlations derived from these parameters were reported to show a more nuanced anisotropic behavior. However, their calculated spin correlation functions display some decoupling of the kagome planes with pronounced correlations along near neighbor [110] directions. 

The nature of the transition near 200mK, and the resulting ground state of Yb$_2$Ti$_2$O$_7$, remains a matter of debate.  Most of the experimental studies indicate a disordered ground state.  This is supported by a recent theoretical study\cite{leonspinwaves} that makes use of the measured high-field spin wave spectrum to determine microscopic exchange parameters. The parameters are highly anisotropic, favoring a local $<$111$>$ exchange.  From a mean field perspective, these parameters predict an ordered ferromagnetic state below 3.2K.   Frustration and quantum fluctuations therefore must be responsible for suppressing a transition to long range FM order, possibly to zero temperature.  Furthermore, the extracted exchange parameters may place Yb$_2$Ti$_2$O$_7$ within the non-magnetic spin liquid ground state studied by Refs. \onlinecite{hermele2004pyrochlore} and \onlinecite{banerjee2008unusual}.  In this scenario, the zero-field QE diffuse scattering measured at 30mK in the most recent single crystal work, Ref. \onlinecite{ross2009}, would represent two-spinon-type magnetic excitations.  

Contrary to this scenario, the work of Yasui \emph{et  al} indicates an ordered, collinear FM ground state below 250mK, albeit possibly a glassy one with a very long relaxation time.\cite{yasui}  In light of the results to be described in the present report, which identify a prevalent sample dependence to the temperature and sharpness of the specific heat anomaly at ``T$_c$'', this FM ordering remains a possibility for the true ground state of Yb$_2$Ti$_2$O$_7$.  However, we will show that the sample dependence of the C$_p$ anomaly is seemingly irrelevant to the changes in energy-integrated diffuse scattering, which occur at a temperature that does not coincide with any specific heat anomaly.  Furthermore, the spread of T$_c$ in various samples is quite small compared to the overall suppression from the predicted mean field transition temperature, lending further support to the idea of a low temperature state dominated by quantum fluctuations. 

\begin{figure}[!htb]  
\centering
\includegraphics[ angle=90,width = 8.5cm]{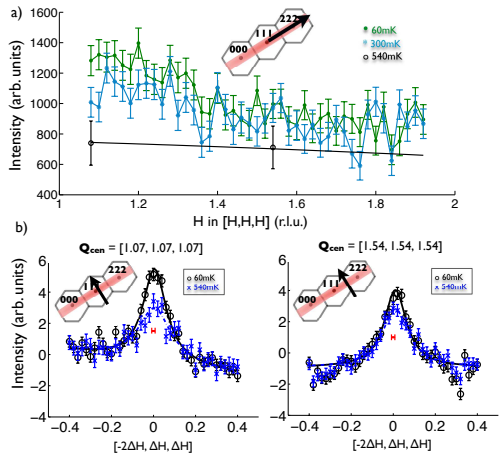}
\caption{ a) (10K background subtracted) Longitudinal rod scans at T = 60 mK and 300 mK.  Data points at 540mK are consistent with a \textbf{Q} dependence that follows the Yb$^{3+}$ magnetic form factor (black solid line). b) Examples of transverse scans across the rod, centered at \textbf{Q}$_{1.07}$ and \textbf{Q}$_{1.54}$.   The insets show the positions of the scans in reciprocal space. }
 \label{fig:fig2}
\end{figure}

This paper details the temperature dependence of the aforementioned changes in diffuse magnetic scattering in zero-field, and determines the extent to which they correlate to features in the specific heat.  Further, through specific heat measurements on various samples of Yb$_2$Ti$_2$O$_7$, we reveal evidence for sample dependence in both single-crystalline and powder samples, which we posit affects all samples reported on in the literature to date.  Through neutron powder diffraction and Rietveld analysis, we show that the sample-dependence is not due to an obvious structural distortion, but rather to small levels of random disorder or strain within an average ordered structure.

\section{\label{sec:level1}Experimental Method}

Single crystalline Yb$_2$Ti$_2$O$_7$ was prepared at McMaster University using a two-mirror floating zone image furnace. It was grown in 4 atm of oxygen, at a rate of 5 mm/h, the growth procedure resembling Ref. \onlinecite{Gardner_growth}.  The single crystal that was the subject of these neutron measurements was the same one studied previously \cite{ross2009}.  The specific heat measurements were performed on this and one other crystal prepared under identical conditions.  The starting material for these growths, a pressed polycrystalline sample, was prepared by mixing Yb$_2$O$_3$ and TiO$_2$ in stoichiometric ratio and annealing at 1200$^\circ$C for 24 hours, with a warming and cooling rate of 100$^\circ$C/h.  The specific heat of this polycrystalline material was measured for comparison to the single crystals.

Single crystal time-of-flight neutron scattering measurements were carried out using the Disk Chopper Spectrometer at the National Institute of Standards and Technology (NIST), with 5\AA \ incident neutrons, giving an energy resolution of 0.09meV.   Triple-axis neutron scattering experiments were also performed at NIST, using SPINS.   For the triple-axis measurements, a pyrolytic graphite (PG) monochromator provided an incident energy of 5meV, and an energy resolution of $\sim$ $\delta E$ = 0.25meV.  Elastically scattered neutrons were selected using five flat PG analyzer blades. Due to the aforementioned energy resolution, the nominal elastic scattering in fact integrates over much of the quasi-elastic component of the diffuse scattering found in Ref. \onlinecite{ross2009}.  A cooled Be filter was used to remove higher order wavelengths from the scattered beam, and was followed by an 80' radial collimator.   

A neutron powder diffraction experiment was performed on two samples of Yb$_2$Ti$_2$O$_7$, using the NDPF instrument at the Lujan Neutron Scattering Center.  One sample studied was a polycrystalline rod (i.e. ``sintered powder''), and the other was a single crystal grown in an identical manner to the sample used in the single crystal neutron experiments.  Both samples were subsequently crushed using a Fritsch Pulverisette 2 mortar grinder, for 20 minutes per sample.  


The heat capacity measurements were performed using the quasi-adiabatic method (see Ref. \onlinecite{quilliam2007specific} for details) with a 1 k$\Omega$ RuO$_2$ thermometer and 10 k$\Omega$ heater mounted directly on the thermally isolated sample.  A weak thermal link to the mixing chamber of a dilution refrigerator was made using Pt-W (92\% Pt, 8\% W) wire for the smaller single crystal piece, A (142.3mg) and polycrystalline sample (24.37 mg), and yellow brass foil for the 7.0515 g crystal, B, used in the neutron scattering experiments.  The time constant of relaxation provided by the weak link was several hours, much longer than the internal relaxation time of the samples, minimizing thermal gradients and ensuring that the sample cooled slowly into an equilibrium state.  The addenda contributed less than 0.1\% to the specific heat of the system.


\section{\label{sec:level1}Results}

\subsection{\label{sec:level2}Diffuse Magnetic Neutron Scattering}

Figure \ref{fig:fig1} (a) shows the temperature dependence of the intensity of elastically scattered neutrons at two \textbf{Q} points using SPINS.  One point is near the fcc zone center and structurally allowed (111) Bragg peak (\textbf{Q}$_{1.07}$ = [1.07, 1.07, 1.07]), while the other is near the fcc Brillouin zone boundary (\textbf{Q}$_{1.54}$ = [1.54, 1.54, 1.54]).  Figure \ref{fig:fig1} (b) locates these two points relative to the rod of diffuse scattering along [111] and the Brillouin zone boundaries, using reciprocal space maps of the elastic scattering at 30 mK and 500 mK measured with DCS.  These two intensities track each other at relatively high temperatures, but they separate below $\sim$ 400 mK, with the scattering near the zone centre (at \textbf{Q}$_{1.07}$) growing more strongly before leveling off at the lowest temperatures.  The inset of Figure \ref{fig:fig1}(a) shows the difference between the elastic intensities at \textbf{Q}$_{1.07}$ and at \textbf{Q}$_{1.54}$, and this resembles conventional order parameter behavior with a phase transition near 400 mK.

\begin{figure}[!tb]  
\centering
\includegraphics[ angle=90,width = 8.5cm]{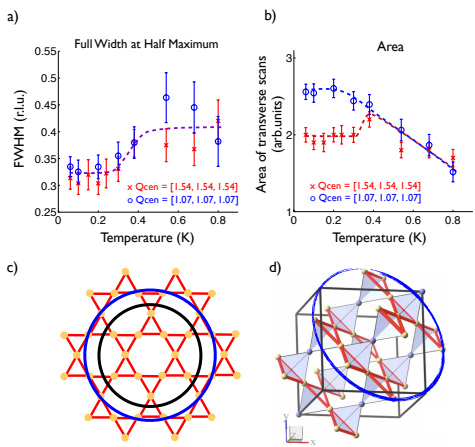}
\caption{ a) Full width at half maxima (FWHM), and b) integrated area extracted from the Lorentzian fits to the transverse scans shown in Fig. 2a). c) The range of $\xi_T$ within the kagome layer.  Above 400mK (black) the correlations are restricted to a single kagome star unit.  Below 265mK (blue), the transverse correlations grow only to encompass the next near neighbors in the kagome layer.  d) The pyrochlore lattice decomposed into kagome and interleaved triangular layers.  Below 400mK, the longitudinal correlations extend only between a single KTK unit. } 
\label{fig:fig3}
\end{figure}

A longitudinal scan of the elastic scattering along the rod (i.e. along the [H,H,H] direction) shows that the low temperature buildup of intensity occurs near the (111), but not (222), zone centre (Figure 2a).   Below 400mK, the scattering peaks up near (111) and the the \textbf{Q} dependence of the longitudinal scattering can be fit to an Ornstein-Zernike (Lorentzian) form, from which an interplane, or `longitudinal', correlation length $\xi_L$ can be determined.  We find $\xi_L$ = 3.2\AA\  at 60mK, implying correlated spins over $\sim$ 6.4\AA \ in the [111] direction.  This indicates that even at the lowest temperatures, 3D correlations do not extend beyond a KTK stack, whose spatial extent is  5.8\AA.  Above 400mK, this interplane correlation vanishes, leaving the Yb$^{3+}$ form factor\cite{formfactor} as the only contribution to the {\textbf Q} dependence of the diffuse scattering along [111] as expected for truly 2D spin correlations.

The scattering normal to the rod probes spin correlations within the kagome and triangular planes.  Figure 2(b) shows scans across the rods centered on \textbf{Q}$_{1.07}$ and  \textbf{Q}$_{1.54}$.  Representative scans at T=60 mK and T=540 mK are shown.  These data sets were fit to Lorentzian lineshapes, allowing the transverse correlation length and integrated area of the scattering to be determined as a function of temperature.  The resolution widths appropriate to the spectrometer are shown as the horizontal bars in Fig. 2(b), and are negligibly small compared to the FWHM of the rod scattering.  The results of the Lorentzian fits are plotted in Fig. 3 (a) and (b).  The integrated areas at the two \textbf{Q} positions differ only for T$<$400mK, (Fig. 3 (b)) consistent with the behavior of the peak intensities along the rod shown in Fig. 1(a).  

Meanwhile, the FWHM (Fig. 3a), which is inversely related to the transverse correlation length $\xi_T$, does not depend on the longitudinal position along the rod at any temperature.   Below 265mK, $\xi_T$ is constant at 9.6\AA\ and it decreases to 7.8\AA\ above 400mK, leaving a correlated area corresponding roughly to one kagome `star', which has a diameter of 15.4\AA.  Combined with $\xi_L$$\sim$ 0 above 400 mK, we conclude that for T$\gtrsim \Theta_{CW}$ the spin system is characterized by uncorrelated kagome stars.  Figure 3(c) summarizes the size of the spin correlated regions in the kagome planes both below 265mK and above 400mK.   In the ground state, which stabilizes below 265mK, short 3D spin correlations form along [111], encompassing a single KTK stack of width $\sim$ 6\AA, as illustrated by the blue region in Fig.3(d).  

\subsection{\label{sec:level2}Quasi-Elastic Neutron Scattering}

\begin{figure}[!tb]  
\centering
\includegraphics[ angle=90, width = 8.5cm]{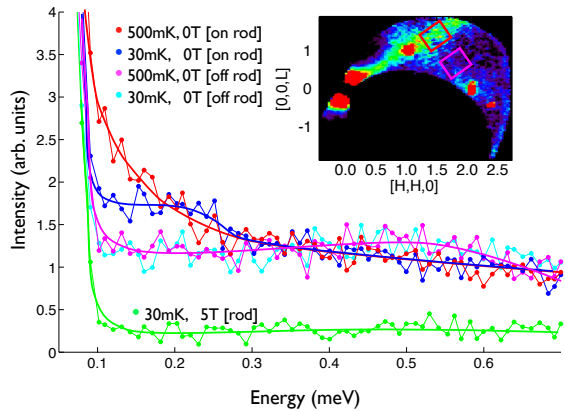}
\caption{Energy dependence of the diffuse scattering above and below T$_c$, both on and off the rod of scattering.   Thick lines are guides to the eye.  Inset: range of binning for the on- and off-rod positions. On-rod: HHH=[1.2,1.6], HH-2H$=[-0.1,0.1]$. Off-rod: HHH=[1.2,1.6], HH-2H=[0.3,0.5]. } 
\label{fig:QE}
\end{figure}

Inelastic magnetic scattering can inform on the dynamics associated with this exotic disordered zero field ground state.  Time-of-flight neutron scattering data is shown in Fig. \ref{fig:QE} at both 30mK and 500mK, and at positions both along the [111] rod of scattering and well removed from the rod of scattering.
These positions are located in the reciprocal space map of the elastic scattering shown in the Fig. \ref{fig:QE} inset.  Consistent with earlier work, this shows the QE scattering along the [111] direction extends to $\sim$ 0.3 meV, but that the spectral weight is depleted at energies less than 0.2 meV, on entering the ground state.  Hence, a dramatic slowing down of the spin fluctuations is observed, consistent with the earlier M\"ossbauer and $\mu$SR results,\cite{hodgesfluc} as the 3D correlated ground state forms.   Application of a [110] magnetic field in excess of $\sim$ 0.5 T, induces a polarized, ordered state, and completely eliminates the diffuse scattering.  Thus, the data in Fig. 3 (c) at 30 mK and 5 T serves as a measure of the background. 

Comparison of the of zero-field quasi-elastic diffuse scattering to the non-magnetic background revealed in a magnetic field of 5T makes it clear that although diffuse scattering is organized into rods of scattering along [111], there is appreciable diffuse, inelastic scattering throughout the Brillouin zone.  This ``off-rod'' scattering has a relatively flat energy dependence out to 0.6meV.  Furthermore, unlike the rod diffuse scattering along [111], the ``off-rod'' diffuse scattering shows no temperature dependence between 500mK and 30mK. This implies that a dynamic and essentially uncorrelated component of the spins persists in the zero-field ground state.  A similar situation is encountered in the classical spin ice Ho$_2$Ti$_2$O$_7$, which retains appreciable \emph{elastic} diffuse scattering throughout the Brillouin zone at very low temperatures, indicating static zero-field spin correlations that encapsulate at most a single tetrahedron.\cite{clancy2009revisiting}

\subsection{\label{sec:level2}Specific Heat}

\begin{figure}[!tb]  
\centering
\includegraphics[width = 8cm]{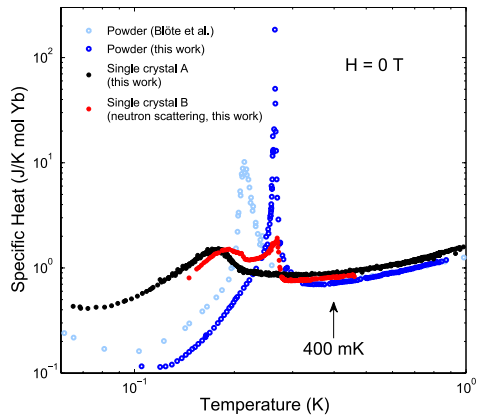}
\caption{ Examples of low temperature specific heat in Yb$_2$Ti$_2$O$_7$. Significant sample dependence is observed in both powders and single-crystals.  The powder sample prepared at McMaster University shows the highest temperature (265mK) and sharpest anomaly.  The neutron scattering sample, B, shows a sharp peak at 265mK similar to the powder, and  a broad, low temperature feature similar to crystal A.  None of the samples studied had any feature at 400mK.}
\label{fig:specificheat}
\end{figure}

The specific heat provides insight into the extent to which the entropy changes at temperatures relevant to the diffuse scattering.  Figure \ref{fig:specificheat} shows specific heat results of poly- and single-crystalline samples, including the sample from the neutron scattering study.  There is significant sample dependence of the low temperature C$_p$ anomaly, with at least one feature occurring between 150mK and 265mK in all samples.  The sharpest anomaly is observed in our powder sample, at a temperature that seems to be the upper limit for all others, T=265mK.  Significantly, our powder C$_p$ peak is sharper by an order of magnitude, and occurs at a higher temperature than both the original powder $C_p$ data by Bl\"ote \emph{et al}, as well as the powder C$_p$ data by Dalmas de R{\'e}otier \emph{et al}, which displays an anomaly at 250mK with a peak height of $\sim$9 J/K mol Yb.\cite{dalmas2006studies} The neutron scattering sample B exhibits a sharp peak at 265mK, but also has a broad feature as seen in sample A.  
  
The specific heat does not show any indication of a feature at 400mK.   This indicates that the buildup of 3D correlations, as well as the small increase in transverse correlation length, produces, at most, subtle changes in the entropy.  This could be another manifestation of the persistently short correlation lengths at all temperatures.  However, an important question remains - what is the physical significance of the  C$_p$ features at 265mK and 200mK observed in the neutron scattering crystal?  Drawing on the large change in spin fluctuation frequency observed by Hodges \emph{et al} at 240mK in their powder sample,\cite{hodgesfluc} we should expect a change in dynamics at these temperatures.  As discussed in relation to Fig. 4, we do observe a change in dynamics between 500mK and 30mK but the intermediate temperature regime awaits further study.

\subsection{\label{sec:level2}Structural Information}
\begin{figure}[!tb]  
\centering
\includegraphics[ angle=90, width = 8cm]{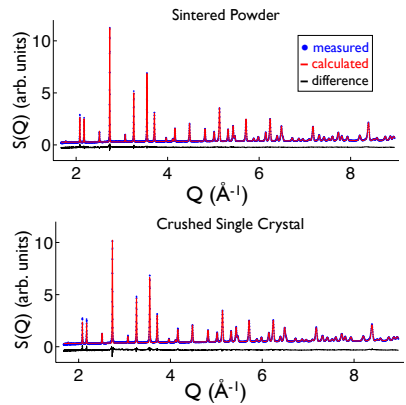}
\caption{Neutron diffraction data from two samples of Yb$_2$Ti$_2$O$_7$ taken at 250K: a sintered powder (top) and a crushed single crystal (bottom).  The result of Rietveld analysis, using a fully ordered model for Yb$_2$Ti$_2$O$_7$, is shown in red.  The difference between the measured and calculated profiles is in black.}  
\label{fig:fig6}
\end{figure}

To investigate structural differences between the single crystals as grown by the method described in Section II, and the polycrystalline material which was used as the starting material for these growths (the ``sintered powder''), we turn to neutron powder diffraction results.  After a successful growth, a single crystal was crushed back into polycrystalline form in order to facilitate direct comparison to the sintered powder. Figure \ref{fig:fig6} shows the two neutron diffraction patterns from the powder and crushed single crystal at 250K.  The data is normalized to the incident time-of-flight spectrum, and corrected for detector efficiency using a vanadium normalization.  A Rietveld refinement was performed on both patterns using a fully ordered model of the pyrochlore structure (see, for example, Ref. \onlinecite{greedan2009local} for details of this structure).  The results indicate that both the powder and crushed crystal are described extremely well by the fully ordered model, with R$_p$ values of 2.16\% and 2.83\% respectively.  Some differences can be observed between the two samples, namely that the crushed crystal has a larger lattice parameter ($a_{xtal} = 10.02006(2)$ compared to $a_{pow} = 10.01322(1)$) as well as greater widths of the Bragg peaks, indicating lattice strain.  Site substitution between Yb$^{3+}$ and Ti$^{4+}$ was found not to be important for either sample (refinement gives less than 1\% substitution in both cases).
A detailed, temperature-dependent Rietveld refinement as well as Pair Distribution Function analysis is forthcoming, but these preliminary results already indicate that there are only subtle structural differences between the two forms of Yb$_2$Ti$_2$O$_7$, pointing to random disorder or lattice strain as the root cause for the differences in specific heat. 

\section{\label{sec:level1}Conclusions}

In summary, we have explored, using diffuse and inelastic neutron scattering,
the transformation of spin correlations as Yb$_2$Ti$_2$O$_7$ enters its exotic ground state at zero field.
Within this ground state, we observe slow, short-range 3D correlated fluctuations that extend over single KTK stacks.
In contrast, at temperatures characteristic of $\Theta_{CW}$ and above, the system is comprised of decoupled kagome stars,
with 2D spin correlations confined to single kagome units.
The temperature T$^{*}$  $\simeq$ 400mK at which correlations lose their 3D nature and also shrink in the 2D kagome plane
is found to be much higher than any of the temperatures T$_c$ $<$ 265mK marking anomalies in C$_p$.
Random structural disorder or lattice strain may account for the variability in T$_c$ and the sharpness of the associated anomalies,
but would seem unlikely to be the factor capping the extremely short range of correlations in the ground state.
Both T$^{*}$ $\simeq$ 400mK and the highest T$_c$ = 265mK observed are an order of magnitude smaller than the predicted mean-field transition temperature, 3.2K.\cite{leonspinwaves}
It is clear that the transition to the exotic ground state, whether this be characterized by T$_c$ and/or T$^{*}$,
is strongly suppressed by geometrical frustration, quantum fluctuations, or both.

The authors acknowledge many useful discussions with M.J.P. Gingras, J.D. Thompson, and P. A. McClarty, and are grateful for technical assistance from Y. Qiu.  This work has benefited from the use of the NPDF beamline at the Lujan Center at Los Alamos Neutron Science Center, funded by the US DOE Office of Basic Energy Sciences. Los Alamos National Laboratory is operated by Los Alamos National Security LLC under DOE contract No. DE-AC52-06NA25396.  This work utilized facilities supported in part by the National Science Foundation under Agreement No. DMR-0944772, and was supported by NSERC of Canada.

\end{document}